# Private Web Browser Forensics: A Case Study of the Epic Privacy Browser

A Reed[1], M Scanlon[2], N-A Le-Khac[2]

[1]*Ottawa Police*
*Ottawa, Canada*

*E-mail: reeda136@gmail.com*

[2]*Forensics and Security Research Group*
*School of Computer Science*
*University College Dublin*
*Dublin, Ireland*

*E-mail: mark.scanlon@ucd.ie; an.lekhac@ucd.ie*

***Abstract:*** *Organised crime, as well as individual criminals, is benefiting from the protection of private browsers provide to those who would carry out illegal activity, such as money laundering, drug trafficking, the online exchange of child-abuse material, etc. The protection afforded to users of the Epic Privacy Browser illustrates these benefits. This browser is currently in use in approximately 180 countries worldwide. This paper outlines the location and type of evidence available through live and post-mortem state analyses of the Epic Privacy Browser. This study identifies the manner in which the browser functions during use, where evidence can be recovered after use, as well as the tools and effective presentation of the recovered material.*

## Introduction

Internet security has been a major and increasing concern for many years, in part because it can be compromised not only through the threat of malware, fraud, system intrusion, or damage, but also through the tracking of Internet activity. In order to combat these threats, encryption of data as a default setting is now commonplace. Firewalls (that is, software that controls access to and from a network) and anti-virus programs are essential tools in the fight against computer crime. Criminals are using numerous methods to access data in the highly lucrative cybercrime business. Organised crime, as well as individual users, is benefiting from the protection of several anti-forensic techniques—including Virtual Private Networks (Conlan 2016), cloud services (Farina *et al*. 2015), and private browsers (Gabet 2016) such as Tor, Ice Dragon, and Epic Privacy Browser—to carry out illegal activity such as money laundering, drug dealing and the trade of child-abuse material (Reed, Scanlon & Le-Khac, 2017). Weak security has been identified and exploited in several high-profile breaches in recent years. Most notably, in 2011, the Sony PlayStation network faced a major security breach (Gazzini& Holt 2011). Over 77 million PlayStation accounts were hacked, which resulted in 12 million unencrypted credit card accounts' being compromised and the site's being closed for a month. In 2005, the United States' Internal Revenue Service (IRS) faced a data breach that resulted in a reported $50 million in fraudulent claims. In 2015, Ashley Madison (Fox-Brewster 2015), a site for extramarital affairs, had 37 million account holders' details released. Breaches such as these underscore the need for better online security and Internet privacy.

Following the Snowden breach (Toxen 2014), there was public outrage at the lack of privacy leading to a rise in the number of browsers offering private browsing. News articles offering advice regarding Internet privacy assisted in educating the public, and a new era of private browsing arose. Although these measures were designed to protect legitimate browsing privacy, they also provided a means to conceal illegal activity. As Rubenking notes, one such tool released for private browsing was the Epic Privacy Browser. This was first released in August 2013 by an India-based company called Hidden Reflex. The Epic Privacy Browser is based on the open-source web browser, Chromium (2014). The Chromium project has resulted in several privacy-enhancing browsers' being built upon its source code, including the Epic Privacy Browser, Comodo (Choi *et al*. 2012), Dooble (Gabet 2016), Inox, and Project Maelstrom (Farina, Kechadi & Scanlon 2015). The Epic Privacy Browser was made available for Windows and OSX operating platforms. As per the browser's homepage, https://www.epicbrowser.com/, it has over one million users and is currently used in approximately 180 countries worldwide (Epic Privacy Browser Homepage, 2017). The Epic Privacy Browser is promoted as a browser specifically engineered to protect users' privacy. It solely operates in private-browser mode and, upon close of the browsing session, deletes all browsing data. Each tab functions as a separate process to increase security. In addition, it claims to remove address bar and URL (Uniform Resource Location) tracking, to remove installation and error tracking, and to offer a 'one-click' option to surf via the company's own encrypted proxy. The intentions of these measures are to hide the user IP address and encrypt all browsing traffic. To prevent searches being indexed per IP address by the search engine providers, automatic proxy routing occurs when the search engines are used.

Information commonly stored on a device using Internet browsers include cache, temporary Internet files, cookie information, search history, passwords, and registry changes. This paper aims to establish what, if any, data relating to the use of the Epic Privacy Browser is produced during the installation and user interaction with the browser. To that end, the authors ran forensic tools such as Process monitor and Regshot (Regshot 2016), captured the live RAM data after use while the system was still running, and examined data acquired post-mortem once the system was shut down. Because of the privacy concerns surrounding Windows 10, it was used as the main platform for analysis. The authors also compared artefacts found on Windows 10 with those available from Windows 7, both set up using default settings and the latest updates. This paper also examines the Epic Privacy Browser's claim that all traces of user activity will be cleared upon close of the application and establishes whether the introduction of Windows 10 has an adverse effect on this claim.

Investigators can use the methods described in this paper to examine a range of Internet-focused Windows applications including, but not limited to instant messaging (Van Dongen 2007; Voorst, Kechadi & Le-Khac 2016), VoIP applications (Sgaras, Kechadi & Le-Khac 2015; Sha, Manesh & El-atty 2016), and peer-to-peer (P2P) network-client applications (Scanlon, Farina & Kechadi 2015; Bissias *et al*. 2016). Experimental results outlined in this paper can also assist researchers who are finding new methods of preserving privacy or aid in the triage process for front-line forensic personnel (Hitchcock, Le-Khac & Scanlon 2016). The contributions of this paper consist of the following items:

- The identification and analysis of Epic Privacy Browser artefact evidence left on Windows 10 and compared with Windows 7 operating systems.
- The outlining of the amount of data recovered from live analysis compared with post-mortem analysis.
- The examination of Epic Privacy Browser artefact evidence unique to Windows 10.
- The identification of forensic tools available to provide effective analysis.

## Background

## Private browsing

Although private browsing has legitimate uses, such as activity on multiple user devices and political restrictions, many individuals are using the shield of anonymity to carry out illegal activity on the Internet. Private browsing is designed in some web browsers to disable browsing history and the web cache. This allows users to browse the Web without storing data on their systems that could be retrieved by investigators. Privacy mode also disables the storage of data in cookies and browsing history databases. This protection is only available to the local device as it is still possible to identify websites visited by associating the IP (Internet Protocol) address at the website.

Aggarwal *et al*. (2010) examined private browsing features introduced by four popular browsers: Internet Explorer, Mozilla Firefox, Google Chrome, and Apple Safari. The authors noted that private browsing modes have two goals: 1) to ensure sites visited while browsing in private leave no trace on the user's computer, and 2) to hide a user's identity from web sites visited by, for example, making it difficult for web sites to link the user's activities in private mode to the user's activities in public mode. The research also identified inconsistencies among the level of privacy afforded to the user when using private mode with the popular browsers and revealed that, although all major browsers support private browsing, the type of privacy provided by each differs greatly. Firefox and Chrome attempt to protect against both web and local attacks while Safari only prevents local issues. In 2013 Marrington *et al.* examined the privacy benefits of the Chrome portable web browser (including private browsing mode) and discovered that browsing traces remained on the host machine after the session ended and the portable storage device had been disconnected.

Plug-ins and extensions being introduced to the browser can change the configuration, render the privacy settings unable to perform as intended, and leave the browser vulnerable to attack. Well known browsers such as Google Chrome, Internet Explorer, Safari, and Mozilla Firefox rely on similar methods to ensure speed and popularity of their product. Web Cache is a popular way of storing data that can be easily and quickly accessed, thereby negating the necessity to find data that has already been used. History databases, thumbnails (small stored images), temporary files, and cookies (user- and site-specific data) all help to speed up the user experience and, in their path, leave a plethora of artefact evidence for examiners to feast on. Many studies have been carried out in this area; and free tools, such as ChromeHistoryView, ChromeCacheView, IECacheView, as well as forensic software such as Internet Evidence Finder, are available to automate the examination process. All the above browsers have the option to operate in private mode.

Research by Khanikekar (2010) indicates that the use of Internet Explorer in 'Protected Mode' runs a 'Low Privilege' process, preventing the application writing to areas of the system that require higher privilege. Hedberg (2013) states that Firefox browser history and search engine keywords are stored in the physical memory of the computer and can still be accessed after the browsing session by way of pagefile.sys or live memory dump. Of particular interest is Google Chrome's 'incognito' mode, as the Epic Privacy Browser is built on top of Chromium. Similar to Firefox, the history, cookies or download lists are not stored on the drive, but held in the physical memory. This still leaves the possibility of pagefile.sys artefact evidence remaining.

## The Epic Privacy Browser

The Epic Browser prides itself on protecting the user's privacy by blocking tracking scripts, creating a new process every time a new tab is opened, and removing installation information amongst other reported features. Forensic analysts have relied on the recovery of Internet artefacts to prove the type of Internet activity as well as to establish the identity of the user behind the keyboard. Epic Browser was released in August 2013, by a company called Hidden Reflex based in Bangalore, India and Washington, D.C. The browser was released in response to increased concerns of Internet activity monitoring by both government and private company interests. It was the first browser built on Chromium that was engineered specifically to protect the privacy of the user. Epic lists, among its many features, the ability to remove all Google tracking as well as to block other companies' tracking attempts. It also offers the option of an encrypted proxy for added security. Rubenking (2014), a journalist with PC Magazine, published a review of the Epic Privacy Browser highlighting some of its main features. Although being powered by the world's leading search engines, Epic is able to prevent data being leaked. The author noted that the browser routes queries through Epic's proxy server automatically, blocking third party cookies and trackers. He also noted that some websites "simply didn't work with Epic".

## Epic Privacy Browser Forensics

This paper will compare the Epic Privacy Browser performance on both the Windows 7 and Windows 10 operating systems. Windows operating systems hold the majority share of the market, with Windows 7 being the most popular at 46.66% of market share, followed by Windows 10 at 13.65%. It is reasonable to conclude, given these statistics, that a forensic examiner is more likely to deal with one of these operating systems than any others, which is why they were chosen for examination in this study. In addition, this research will establish whether the introduction of new data collection methods presented in Windows 10 have provided an opportunity for forensic investigators to utilise any potential breaches in Epic's privacy settings; whether tools currently used for the analysis of similar browsers built on the same source code, such as Google Chrome, can also be used to recover data from Epic; and whether live analysis, by the capture on Random Access Memory data, differs when using Windows 10 compared to Windows 7.

For the analysis of the Epic Privacy Browser on both the Windows 7 and Window 10 operating systems, a 320GB hard drive was used in an HP desktop computer containing 4GB of RAM. The hard drive was wiped, using Wipemaster hardware, according to Department of Defence standards. Windows 7 Pro was then installed on the hard drive, and all default settings were selected. The computer tower was then connected to the Internet via an Ethernet cable, and all available software and security updates were carried out. Standard firewall and defender settings were applied.

Once the Windows software was updated, the Epic Privacy Browser was installed. Installation of the browser was monitored using the following software to analyse activity on the system:

- Process Monitor – an advanced monitoring tool that shows real-time file system, registry, and process thread activity;
- Regshot – an open-source utility that allows snapshots to be taken pre- and post-software installation in order to record registry changes on the system;
- TCPView – a tool that shows detailed listings of all TCP (Transmission control Protocol) and UDP (User Datagram Protocol) endpoints as well as network connection status;
- Registry viewer – software that allows analysis of the windows registry system;
- FTK Imager – forensics software that is used to capture RAM dumps and protected files data on a live system;
- WireShark – Network protocol analyser that identifies all network traffic;
- ChromeHistoryView – freeware that allows an examiner to view History database records;
- ChromeCacheView – freeware that allows the examiner to view cache entries.

Following installation of Epic, a series of functions were carried out and recorded for the examination. These included Internet searches; viewing of photos, videos and galleries; as well as document and image downloads. Social networking sites such as Facebook, Twitter, Instagram and YouTube were visited. Any login details were entered; and, when offered, the passwords were stored. Google's Gmail was also visited, and account sign in and log out completed. The computer was constantly connected to the Internet for a period of three days with the Epic Privacy Browser displayed. On closure of the browser, but while the computer was still running, the Random Access Memory data was then acquired using FTK Imager (version 3.1.1.8). Protected files such as registry Sam, System, Security, Software, and User files such as NTUSER.DAT was also acquired using FTK Imager at different stages of the process. Upon completion, the system was powered down using the Start>Power>Shutdown option. The same 320GB hard drive was then wiped (again to Department of Defence standards) and placed back into the HP tower, and the process was repeated but this time using Windows 10 Pro operating system with the same browser and forensics software installed. The same queries that had been performed with Windows 7 were repeated. Random Access Memory data was captured before the Epic Privacy Browser was installed and on completion of the search queries, while the browser was still displayed. On completion, the browser was closed and the system shut down using the Start – Power – Shutdown method.

## Live-memory acquisition

As memory capture and analysis become better understood, improved forensic tools have been developed to assist investigators in extracting and interpreting this data. Traditionally, memory analysis has often been avoided due to the complicated nature of acquisition and interpretation, but with the advent of software such as FTK Imager, OS Triage, and Belkasoft RAM Capturer, these processes have become more straightforward. Software features such as improved GUIs, 'push button' applications, and built-in detection functions have made memory retrieval and analysis far less intimidating for the forensic examiner. A great deal of information can be gained from live memory analysis, making live data capture more important than ever before. Information such as network connections and malware communication (often used as a defence) can be established or eliminated through RAM analysis. User names and passwords, as well as decrypted programs, may be found and private browsers, such as Epic, often use RAM in preference to other forms of storage. For these reasons, the authors chose to use live-memory capture as an adopted approach for this study.

In these experiments, a 128GB Thumb drive was used as storage for the RAM and protected file dumps. FTK Imager forensics software was installed on the examination computer on initial set up and was the software used to extract both the RAM and protected file data. The resulting data dump was then transferred to a forensics workstation and labelled as either Windows 10 or Windows 7, pre or post examination, and protected file dumps.

## Post-mortem data acquisition

Computer examiners often receive a device post mortem, meaning that the device has been powered down or the power plug has been pulled, thereby clearing all of the RAM data. The benefits of powering down a device include isolation from a network, prevention of a wipe command deleting the data, and the ability to carry out the search and seizure of equipment without the need for an on-scene computer examiner. In addition, sometimes a device is submitted for examination months after its seizure, and, even then, analysis of the data may not immediately follow. So keeping the device powered on is not always practical or feasible. Given these occurrences, post-mortem data examination was also conducted.

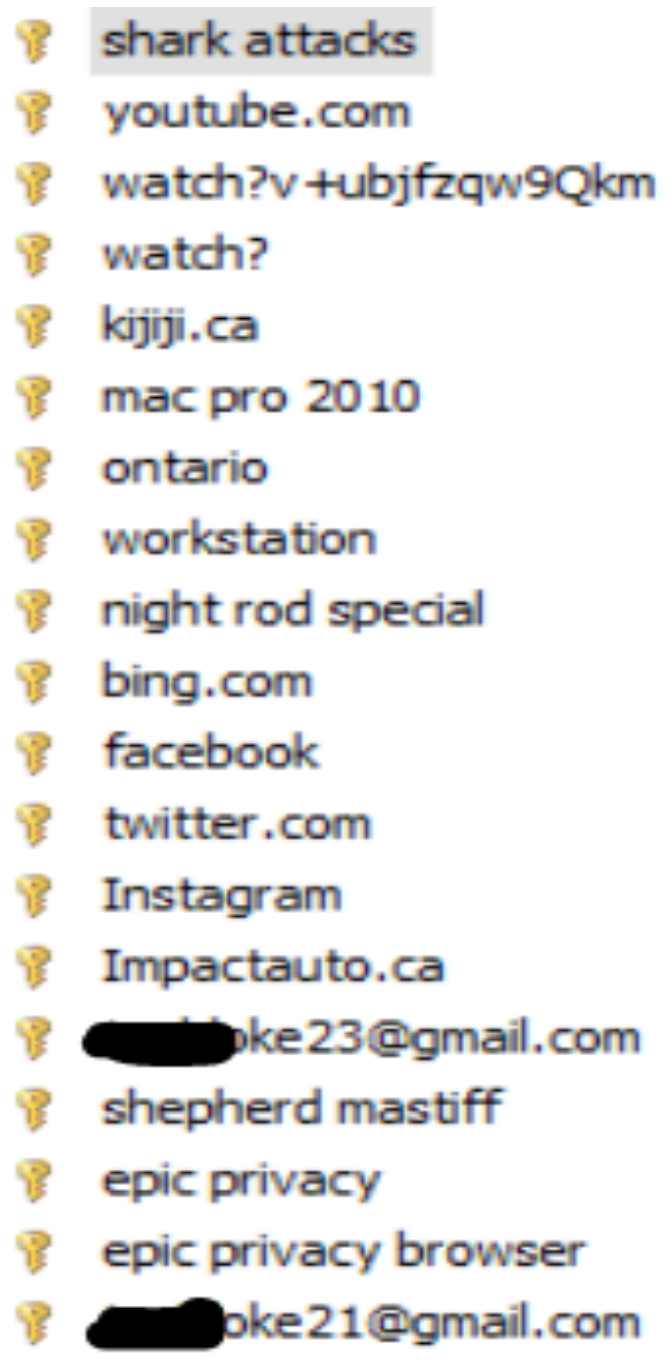

**Figure 1:** List of keyword search terms

In this study, once each hard drive was removed from the HP Tower, it was acquired individually using FTK Imager forensics software via Tableau Write Blocking hardware. This method is used in order to ensure an exact forensics image is obtained and verified by way of Cyclic Redundancy Check, an error-detecting code that detects changes to raw data, and Hash MD5 algorithm on completion of the process. Tableau Write Blocking hardware is connected directly between the hard drive being acquired and the forensics computer running the acquisition software. Its function is to allow read-only commands to be sent to the hard drive, thereby preserving the original data. As the original hard drive is the best evidence in a case required for court, an exact forensic copy is produced as a 'working copy' for investigators to analyse to minimise the risk of damage or data loss to the original hard drive.

Both Windows 7 and Windows 10 E01 files were loaded into Encase forensics software for analysis (version 6.19.7). A 'lost folder' recovery was then carried out followed by the inclusion of the live-memory data. The authors then carried out a search on a number of keyword search terms, as seen in **Figure 1**.

## Windows 7: Epic Privacy Browser Forensic Analysis

## Post-mortem analysis

Initial analysis was carried out on the Epic Privacy Browser installed on Windows 7 professional. The installation was monitored using Regshot freeware. A capture was taken before and after install. The software then compares the before and after snapshots and provides a report of the changes recorded in the registry. Of interest to an examiner would be the application path as well as the version number: C:\Users\User\AppData\Local\Epic Privacy Browser\Application\39.0.2171.71. This contain this default folders, as shown in **Figure 2**. On executing the browser, several other folders and files are created.

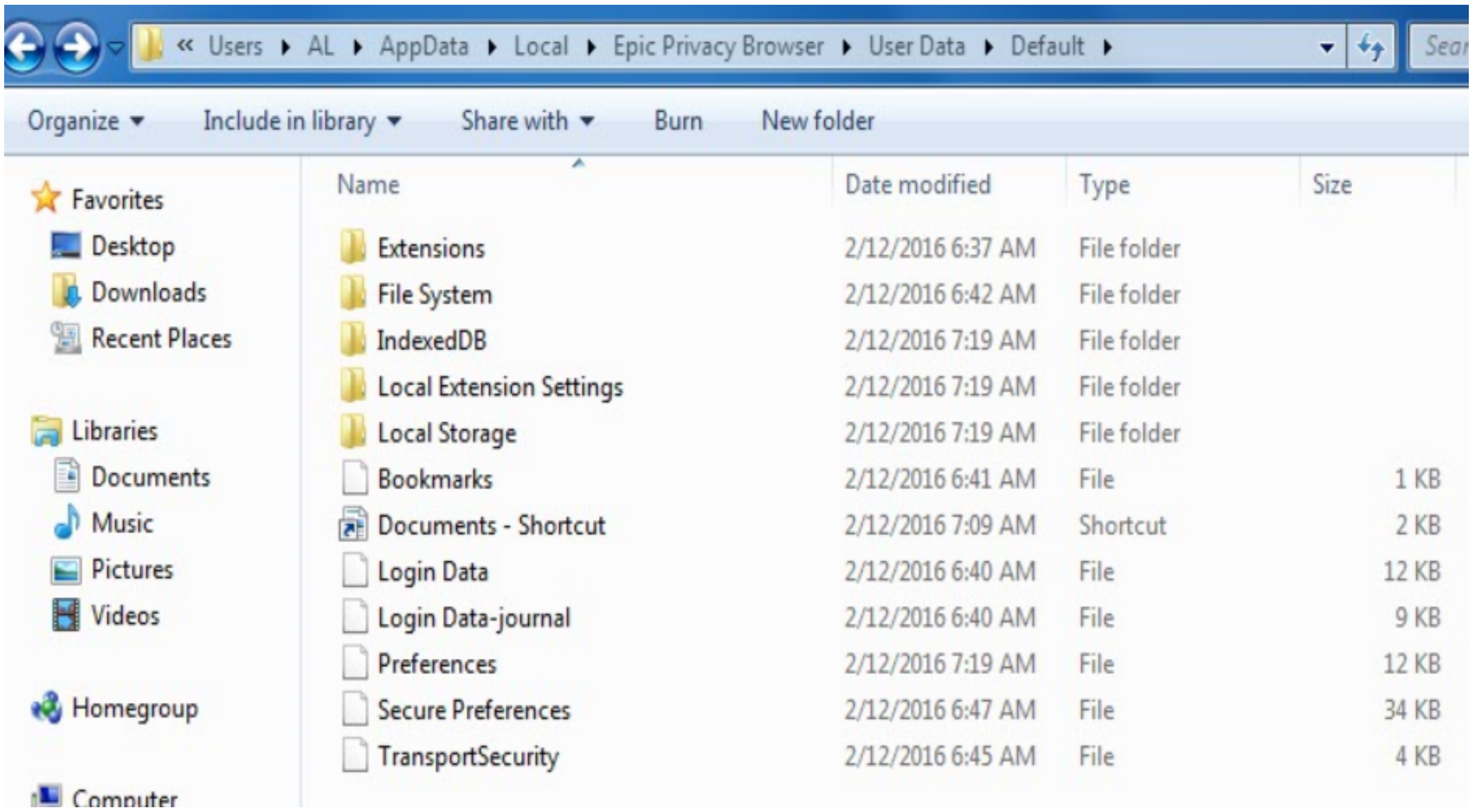

**Figure 2:** Epic default folder contents on install

The folder structure has a very similar look to that of Google Chrome, shown in **Figure 3**.

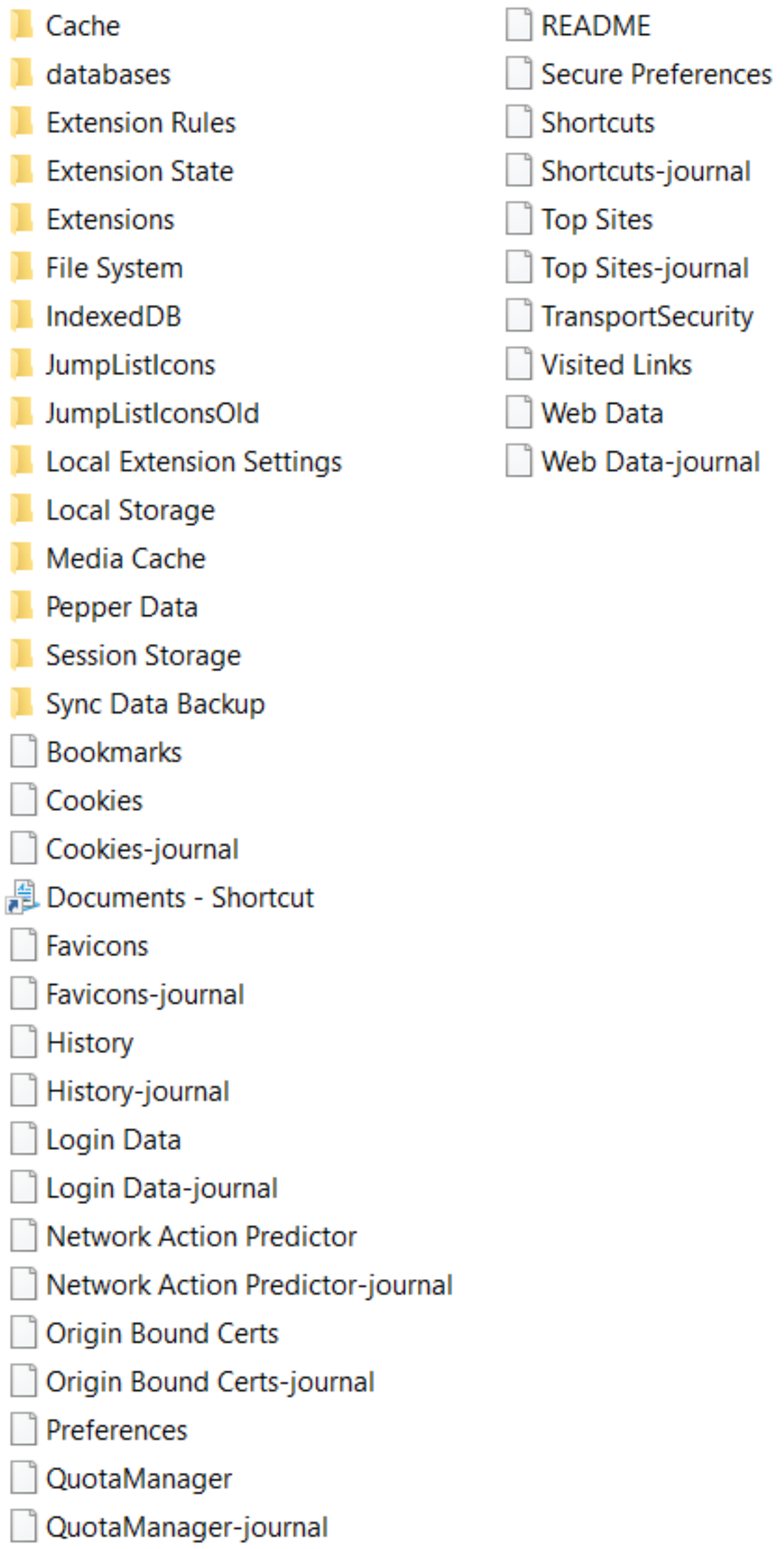

**Figure 3:** Epic default folder on execution

Process Monitor software was used to analyse the browser application launch, **Figure 4**, below. This figure shows the browser making use of a cache folder and additional files that were not initially present on the browser install.

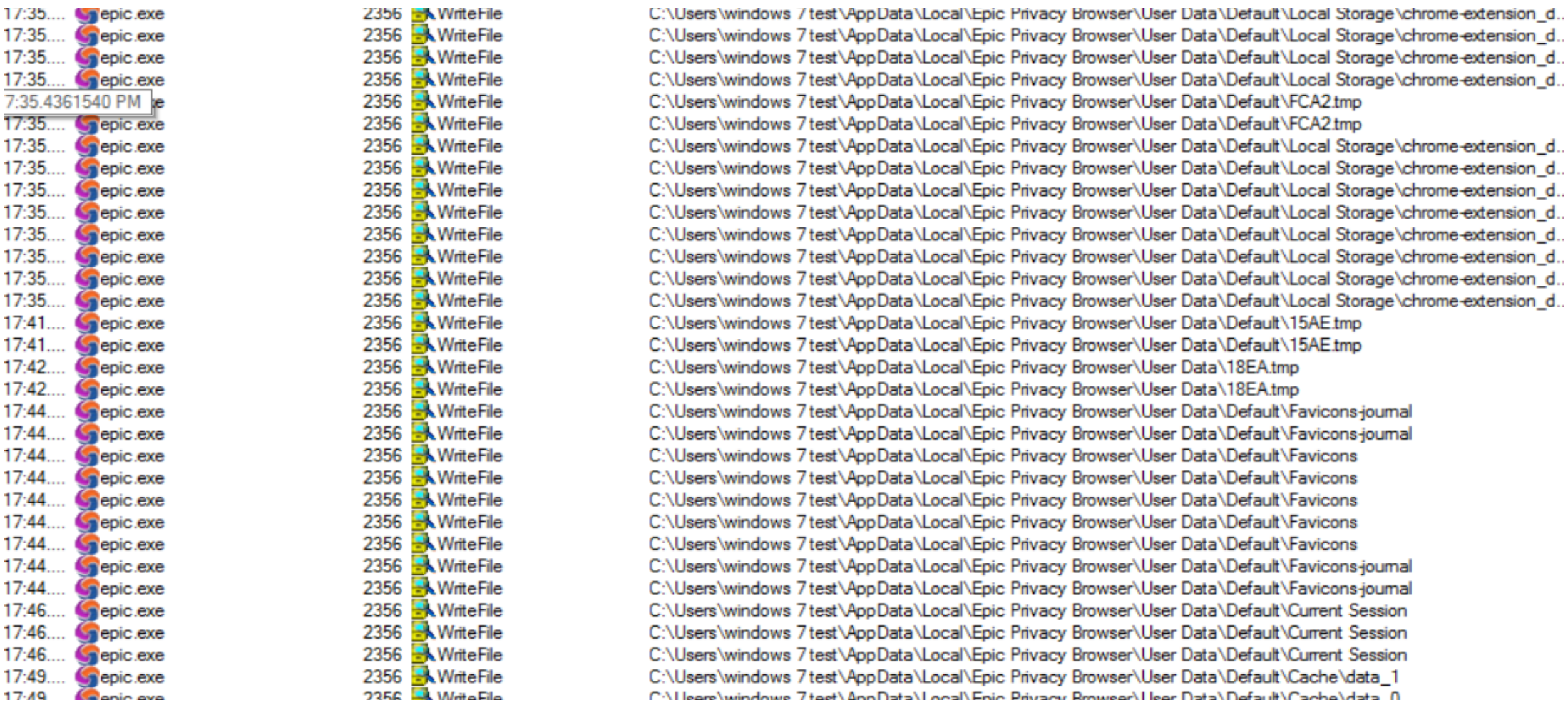

**Figure 4:** Procmon capture of Epic Privacy Browser launch

The additional files and folders are populated with data while the browser is running and deleted when the browser is closed. The *history.db* file and cache folder appear to function in the same way as Google Chrome, allowing data to be viewed using standard Chrome freeware tools. Through the device's live memory capture, tracks of the browser running are recoverable after the browser's closure. Although a large number of files are deleted from view when the browser is closed, a great deal of artefact evidence was either written to pagefile.sys, shown as deleted but recovered using standard forensics tools or recovered from unallocated space. Encase, as well as Internet Evidence Finder, was also able to recover created dates from Epic files shown as deleted, as can be seen in **Figure 5**.

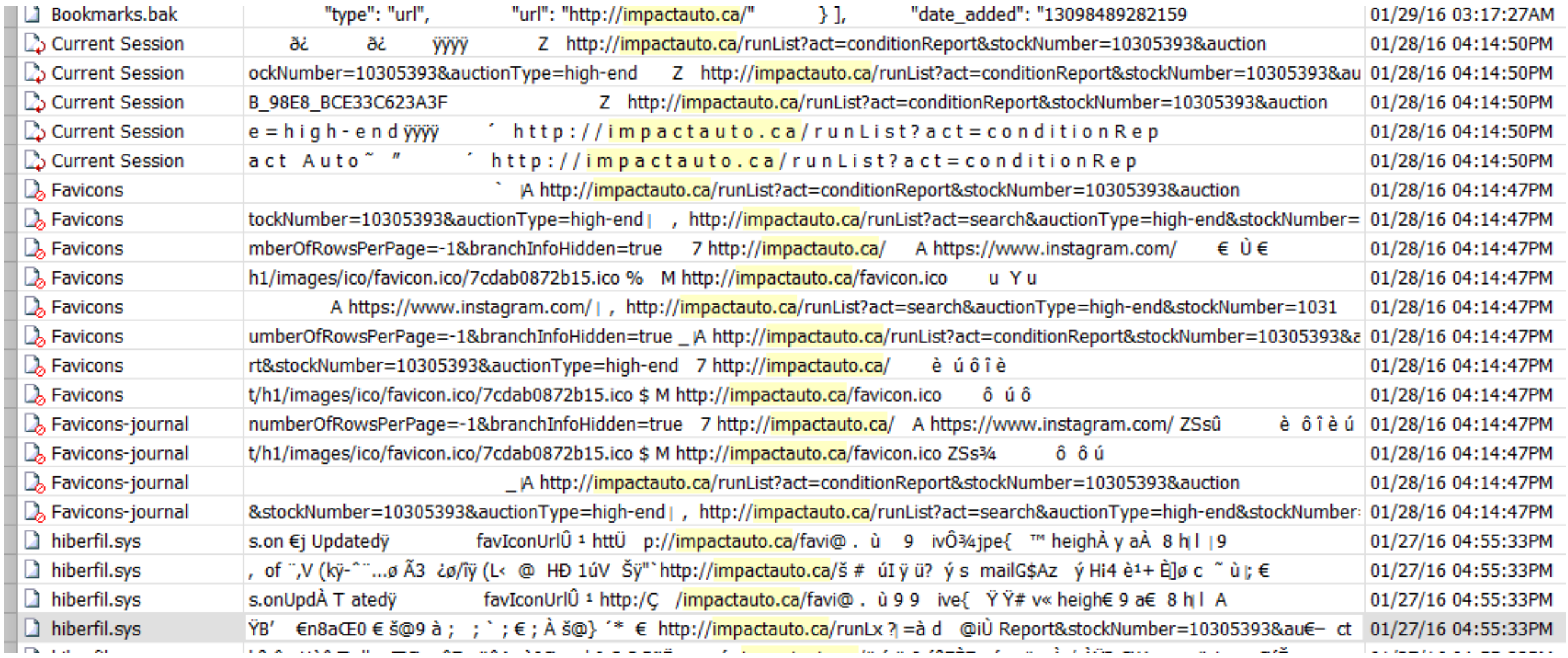

**Figure 5:** Encase screenshot of recovered Epic artefacts including created dates

Further analysis was carried out using Internet Evidence Finder, version 6.6.3.0740. The software allows for the Windows 7 image file to be loaded and specific category searches selected, as can be seen in **Figure 6**.

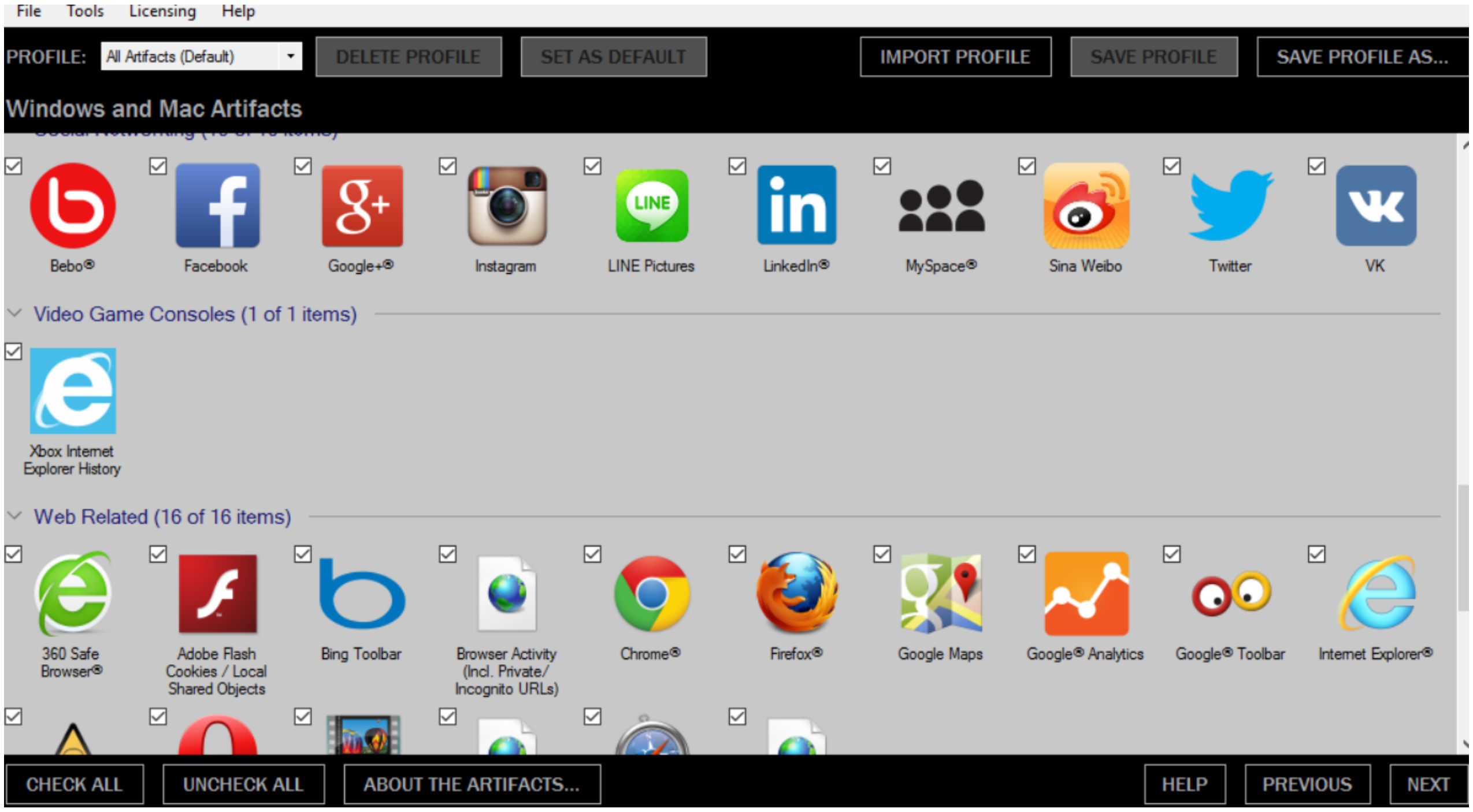

**Figure 6:** IEF software data selection GUI

IEF identified a large number of hits relating to queries carried out during the experiment. It appeared that data was regularly captured and transferred to the pagefile.sys. **Figures 7** and **8**, below, illustrate a small sample of those found.

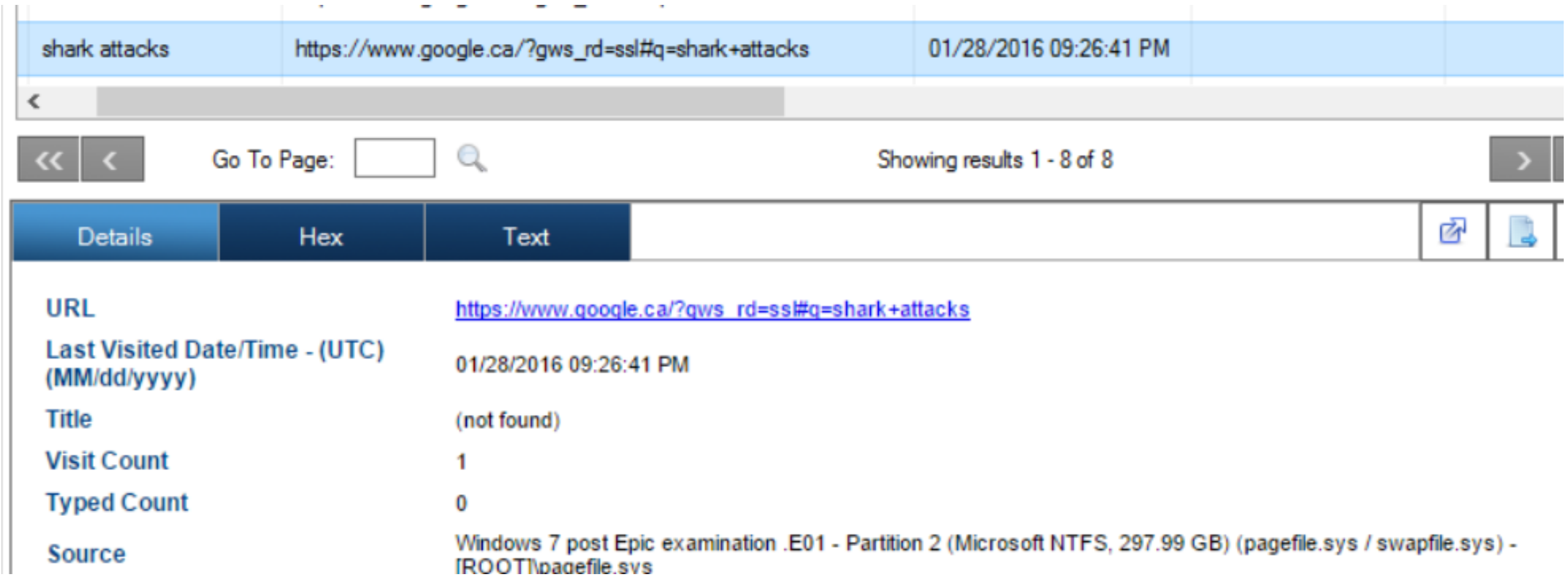

**Figure 7:** Google Search within Epic – ‘shark attacks’ date and time stamped

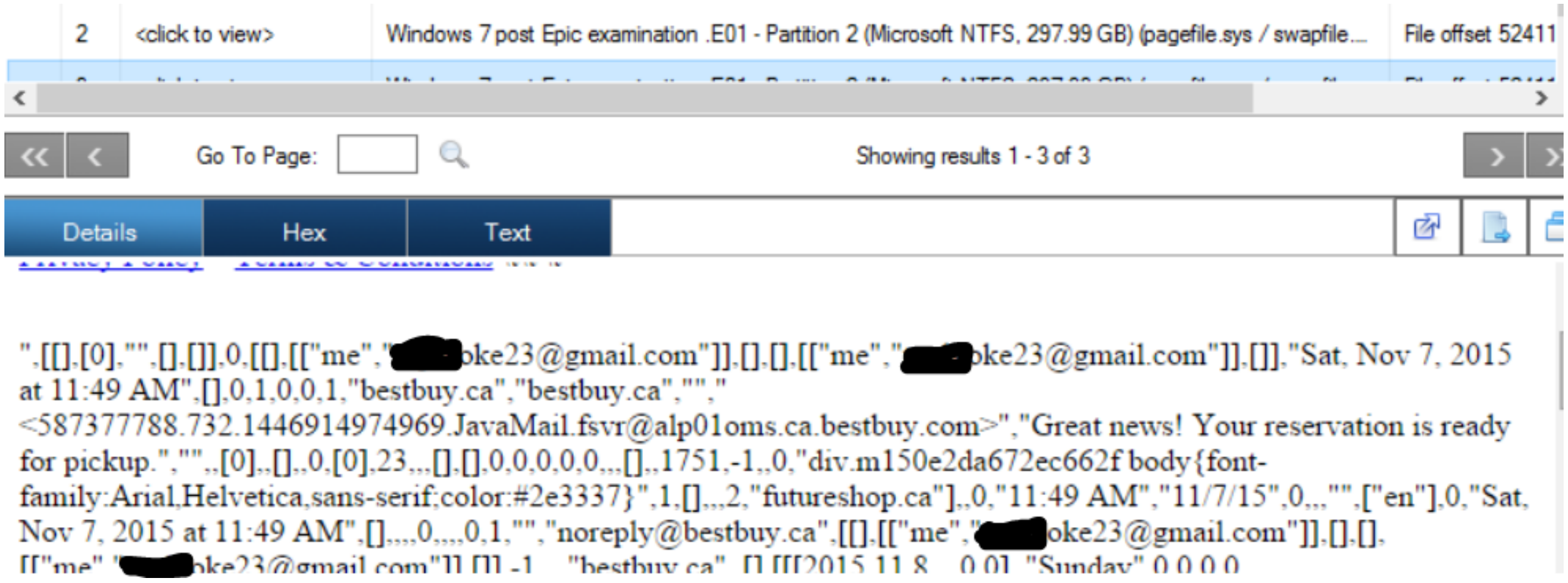

**Figure 8:** Gmail account details, captured in *pagefile.sys*

Windows 7 drive image (E01) returned 343,000 hits from keyword searches (see **Figure 9**). The same keyword search terms were run on the Windows 10 drive image, resulting in only 52,000 hits.

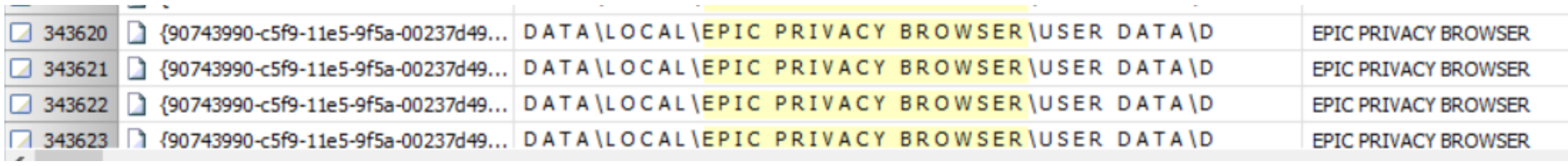

**Figure 9:** Windows 7 drive image keyword hits

## Live analysis

On the completion of the Internet queries, but before the Epic Privacy Browser was closed, the live-memory capture was carried out using FTK Imager software. The system files were also captured at this time. The extracted data was then analysed in both Encase and Internet Evidence Finder. The benefit of live-data capture was immediately evident although, in this case, post-mortem analysis had also bore significant fruit. It appeared that Epic Browser activity on Windows 7 was being both captured in RAM and written to pagefile.sys. Internet Evidence Finder was an excellent tool for parsing out and presenting the evidence found. Of note were the areas that would be beneficial to a forensics examiner (see **Figure 10**).

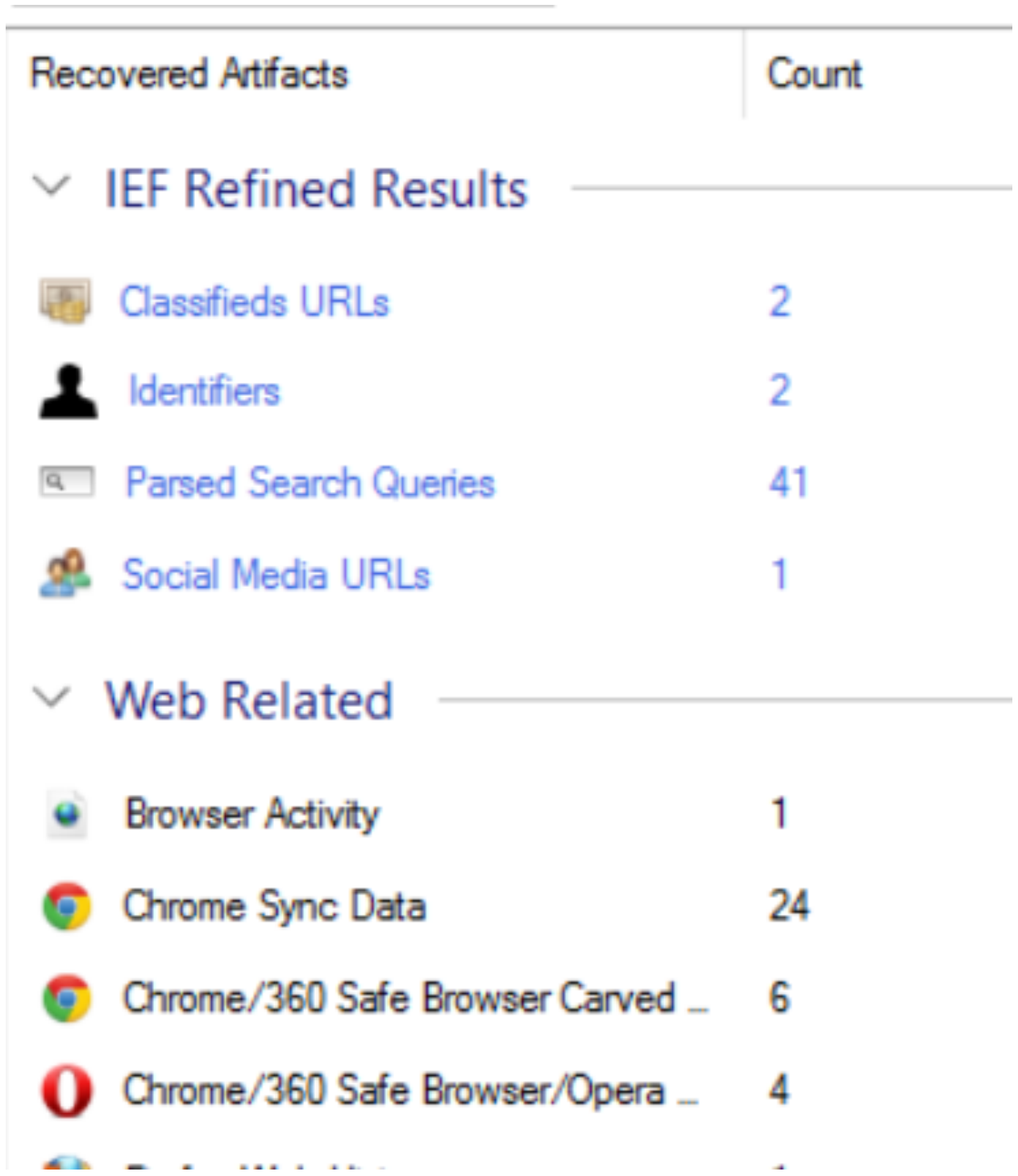

**Figure 10:** IEF Windows 7 total hits on RAM dump

Indeed, **Figure 11**, below, shows the 'kijiji dogs' selection made during the browser query process. This information was retrieved from both the RAM and post-mortem data dumps with the date and time of the search clearly visible.

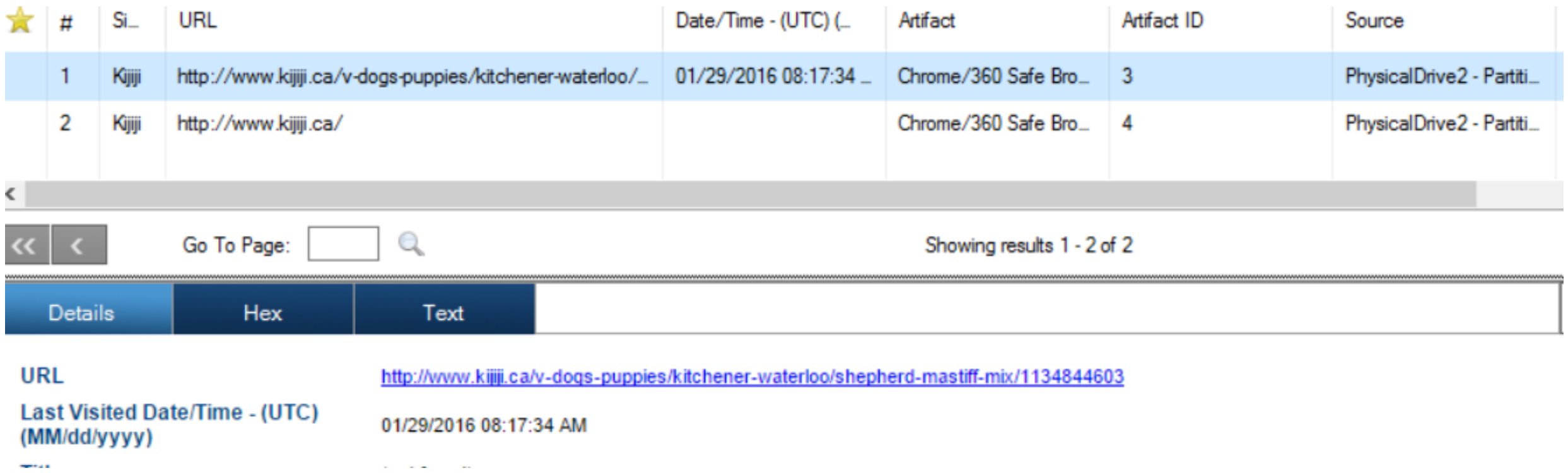

**Figure 11:** Kijiji search for dogs

**Figure 12**, shows a list of the URLs visited during the query stage. The URL "https://epicsearch.in/search?pno=1&q=kijiji" indicates not only the use of Epic, but also that a "kijiji" search was carried out by the user.

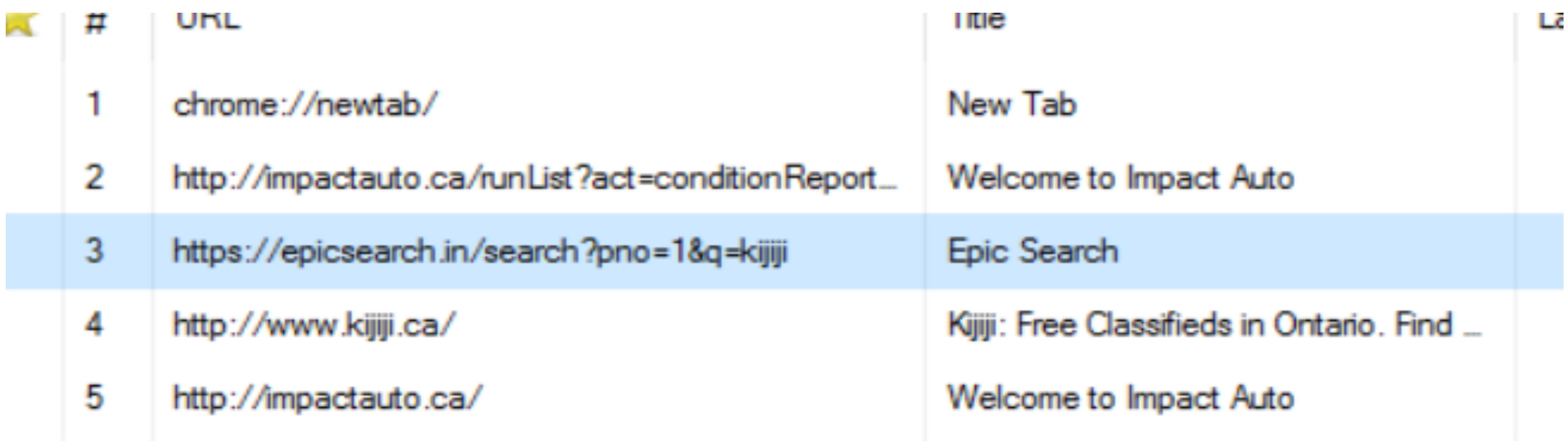

**Figure 12:** URLs visited

IEF returned over 40 hits of interest from the Windows 7 RAM dump, cementing the requirement for investigators to capture live memory when possible, as shown in **Figure 13**.

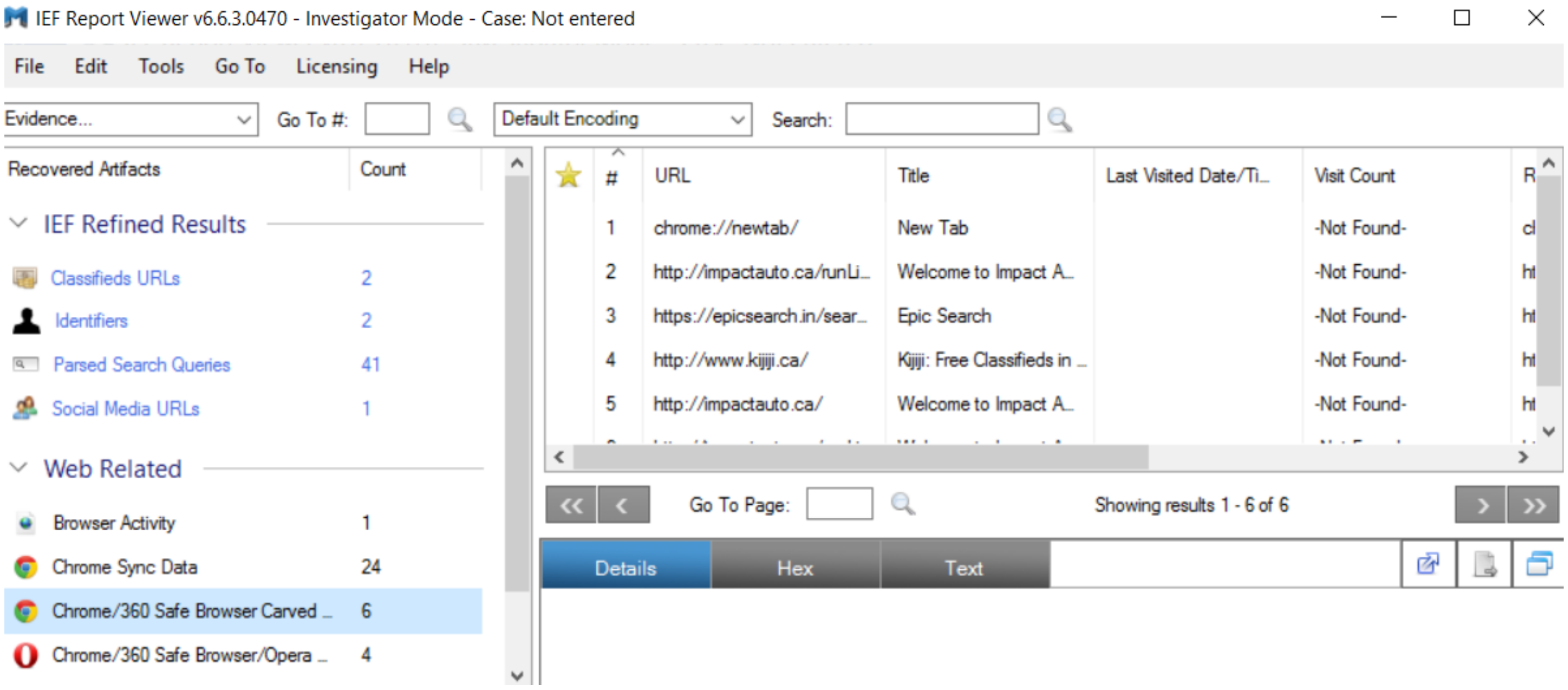

**Figure 13:** IEF total Windows 7 RAM dump hits

Fifty-two thousand hits were recorded from the combined keyword searches entered in Encase, against the live memory dumps of Epic queries on both Windows 7 and Windows 10 operating systems. Of the 52,000 hits, only 12,000 were recorded from the Windows 7 operating system, even though the same experimental process was carried out on each operating system.

## Windows 10: Epic Privacy Browser Forensic Analysis

## Post-mortem analysis

As with Windows 7, the Epic Privacy Browser installation on Windows 10 Professional was monitored using Regshot and Process Monitor tools. A snapshot was also taken immediately before, and after, the installation process to identify changes to both the file system and Windows registry. There were a number of registry entries of interest that had not been present in the Windows 7 install (see **Figures 14** and **15**).

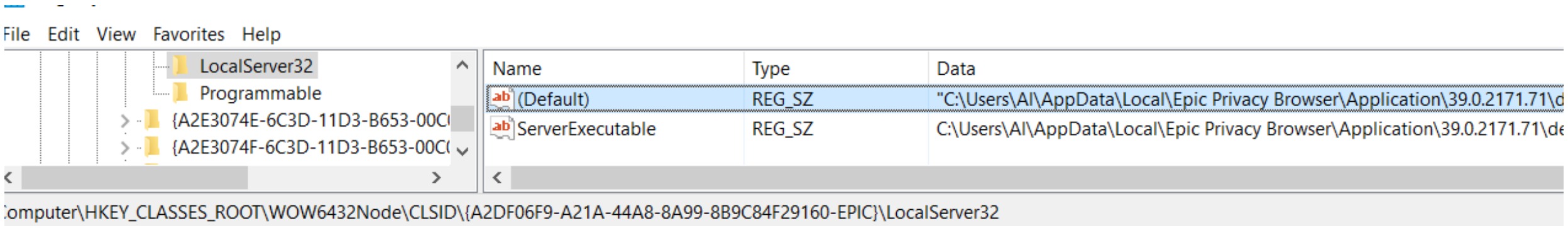

**Figure 14:** Epic WOW6432 node version #

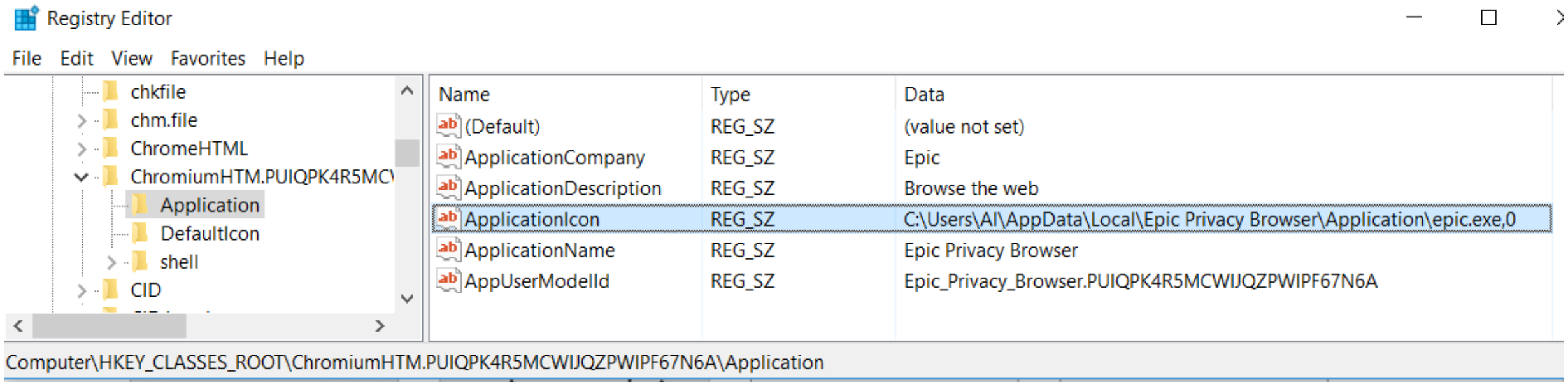

**Figure 15:** Classes root entry

Further entries were discovered specific to the user's Security Identifier (SID) that would assist the examiner in identifying the user account associated with the application. The SID is a device and account identifier. It is variable in length and encapsulates the hierarchical notion of issuer and identifier. It consists of a 6-byte identifier authority field that is followed by 1-14, 32-bit sub-authority value. It ends in a single 32-bit Relative Identifier (RID). This not only makes it unique to the user but also to the device. The SID is assigned during the installation of the operating system and is unique to each computer. All user accounts are based on the computer's SID and contain the relative identifier for each user account. Although this is randomly generated, it is theoretically impossible for the same SID to appear on two devices and is, therefore, extremely useful to a forensic examiner, as can be seen in **Figure 16**.

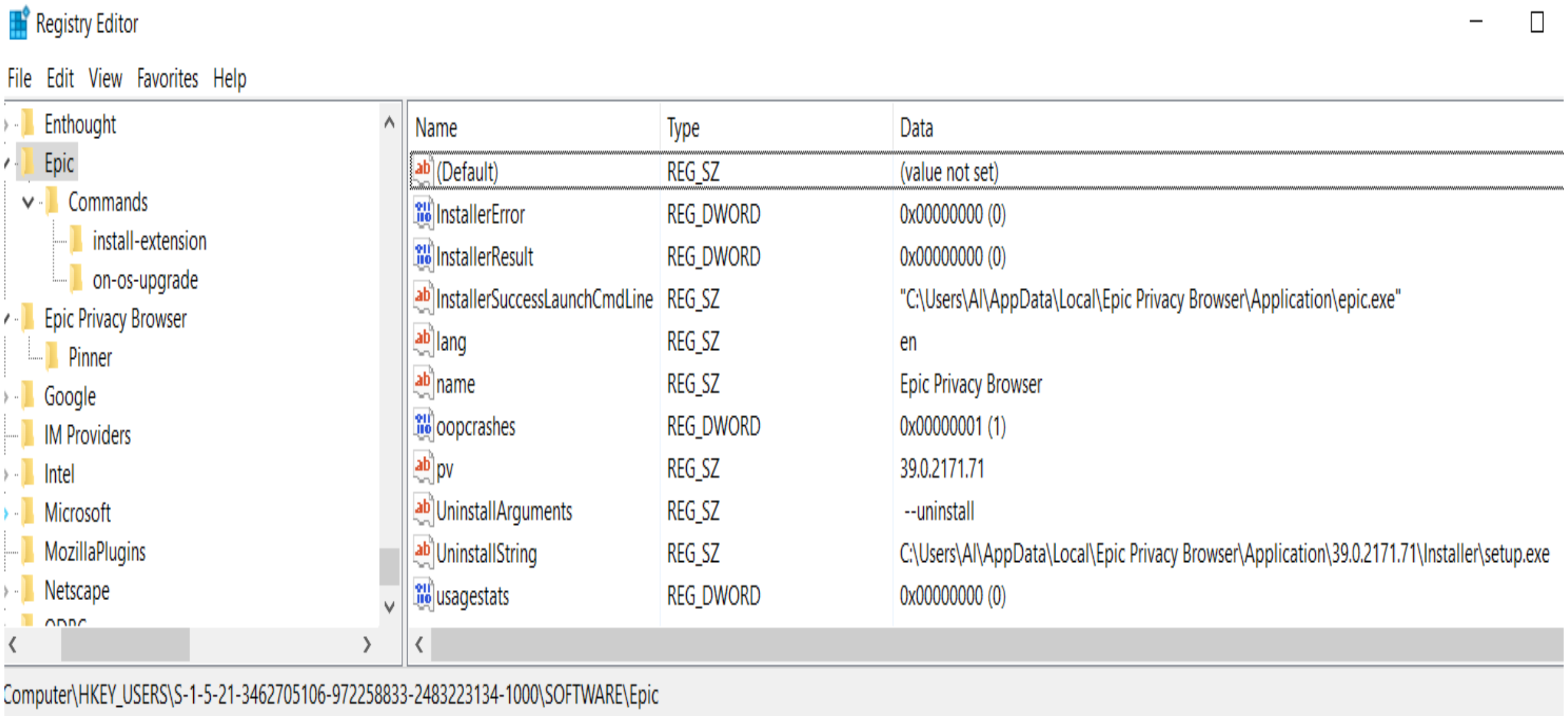

(a) Epic Install SID information

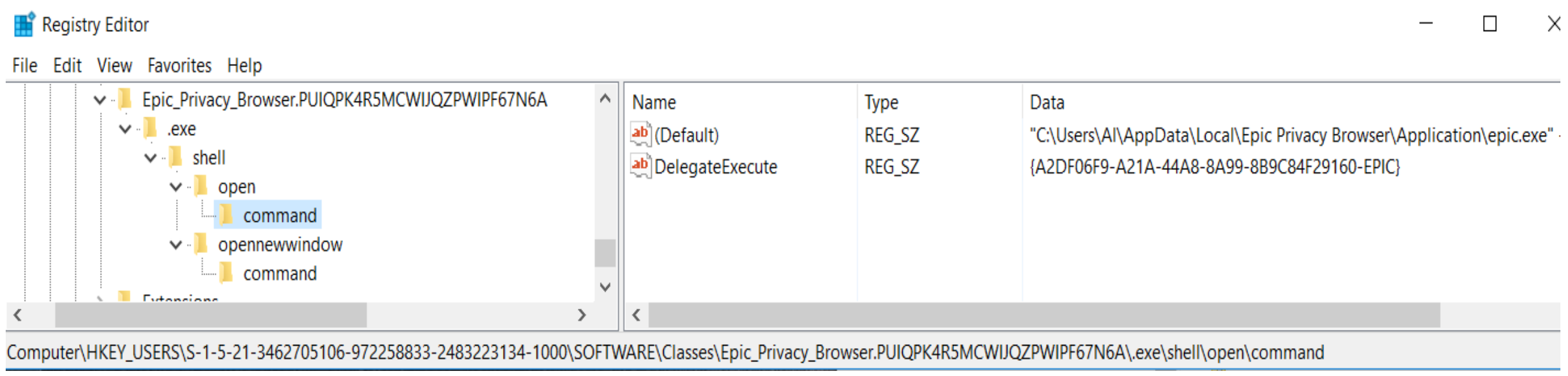

(b) SID 1000 command entry

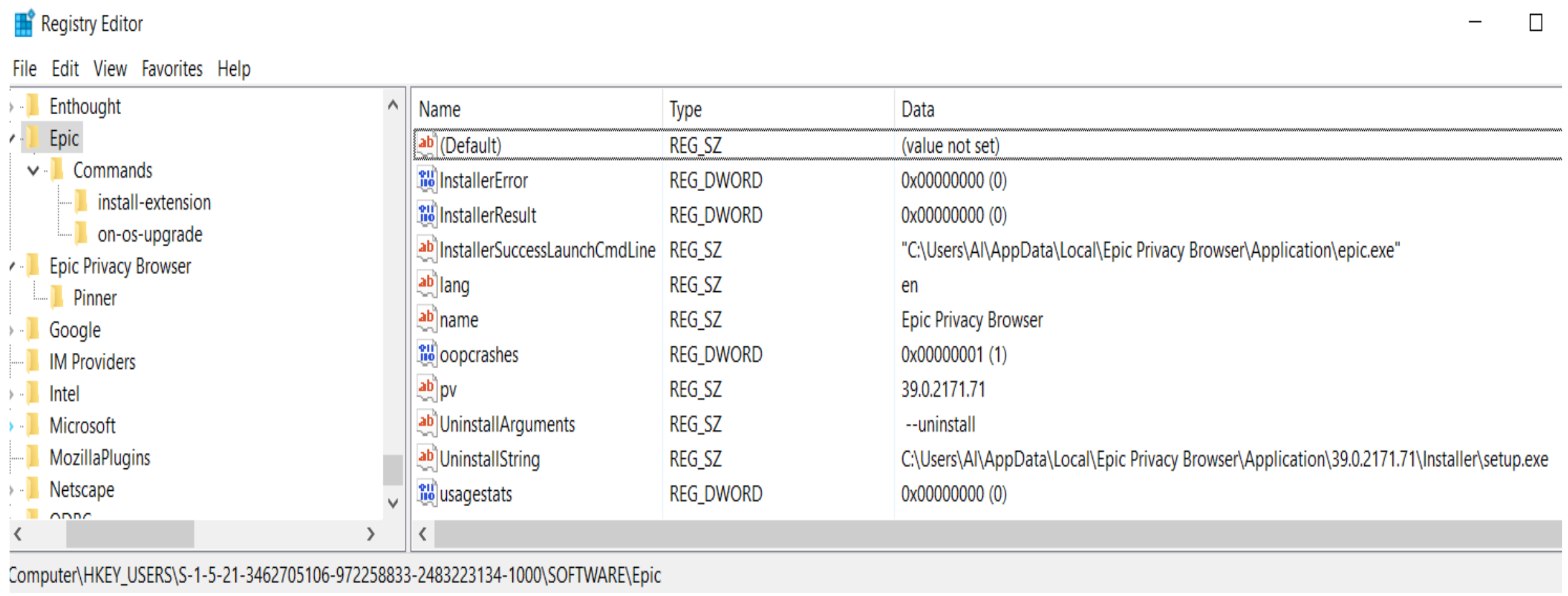

(c) Epic Software entry in 1000 SID

**Figure 16:** SID information

The installation of the Epic Privacy Browser on Windows 10 appears to differ slightly from that of Windows 7 with the addition of a 'Bookmarks.bak' file. This appears to be a backup of the bookmarks file and remains, even when the browser is closed. All other files appear to behave in the same way as in Windows 7 in that the additional cache folder and files are generated on the launch of the browser and then are deleted immediately on its conclusion. A running system with browser displayed offers the best opportunity to capture the default folder and, therefore, the complete history and cache but all is not lost if the system is powered off. Although many of the files display in Encase as deleted, the data, and often the metadata, appears to be present (see **Figure 17**).

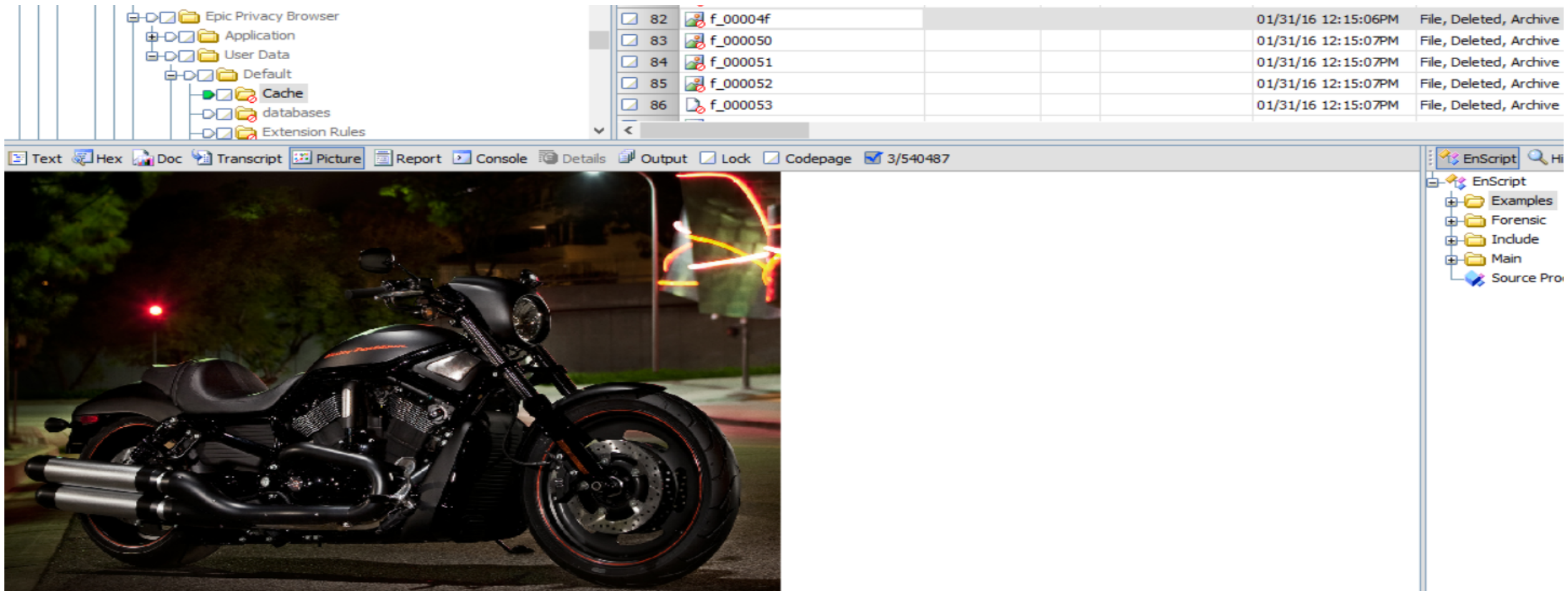

**Figure 17:** Night rod special cache shown as deleted data

Stored in Windows\ServiceProfiles\NetworkService\ is a file named NTUSER.DAT.LOG2 (**Figure 18**).

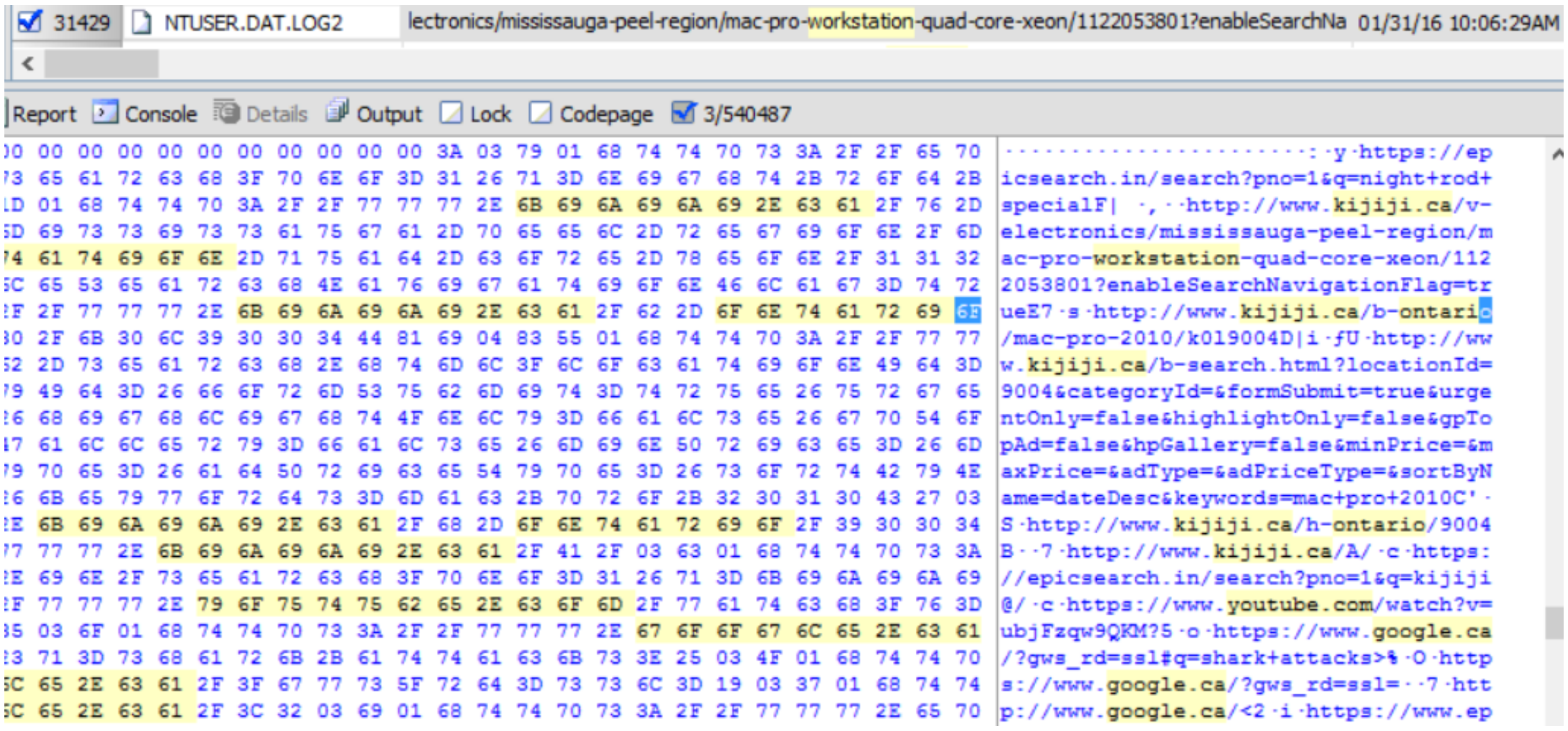

**Figure 18:** NTUSER.DAT.LOG2 file information

The file logged search queries carried out during the experiment, including the site contacted to carry out the search. https://epicsearch.in. The Windows 10 drive image (E01) returned 52,000 hits from keyword searches. The same search terms resulted in 343,000 hits on the Windows 7 image. It appears that live capture for Epic artefact evidence in Windows 10 is far more beneficial compared to Windows 7.

## Live analysis

Windows 10 relied heavily on live-memory storage during the use of the Epic Privacy Browser with the analysis reporting that the newest offering from Microsoft was responsible for approximately 80% of the live captured data compared with the same tests on Windows 7, again enforcing the importance of live-data capture. Encase and IEF were used to analyse and present the data. IEF results of note are illustrated in **Figure 19**)

| | | | | |
|---|---|---|---|---|
| 1 | https://www.youtube… | 01/31/2016 05:10:38 … | 5 Horrific Shark Attack… | 2 |
| 2 | https://www.google.c… | 01/31/2016 05:08:46 … | | 1 |
| 3 | http://google.ca/ | 01/31/2016 05:08:31 … | Google | 1 |
| 4 | https://www.google.c… | 01/31/2016 05:08:31 … | Google | 1 |
| 5 | http://www.google.ca/ | 01/31/2016 05:08:31 … | Google | 1 |
| 6 | https://www.epicbrow… | 01/31/2016 04:54:49 … | Epic Privacy Browser | 1 |
| 7 | http://windows.micros… | 01/31/2016 11:00:45 … | Downloads - Microsoft … | 0 |

Go To Page: Showing results 1 - 94 c

**Figure 19:** Search results captured in live memory

Data within the Gmail account, that was displayed but not directly accessed, was also captured in memory and parsed by IEF. Information of this nature is invaluable to any forensic investigator as it is often difficult to place a user behind the keyboard (see **Figure 20**, below).

| # | Email(s) | Subject | Sent Date/Time - Local Time | Last Modified or Viewed D. |
|---|---|---|---|---|
| 1 | auctions <auctions@i… | IAA Watch List Vehicle(s) now scheduled for sale | Thu, Oct 8, 2015 at 6:55 AM | 11/10/2015 01:00:29 AM |
| 2 | auctions <auctions@i… | IAA Watch List Vehicle(s) now scheduled for sale | Wed, Oct 7, 2015 at 12:15 PM | 10/07/2015 05:07:34 PM |
| 3 | eRepairables.com <inf… | BMW X5 Selling on eRepairables.com | Sun, Jan 31, 2016 at 12:00 PM | 01/31/2016 05:01:01 PM |
| 4 | Air Canada Flight Offer… | Florida sale, summer in Europe, and cruise savings! | Wed, Jan 13, 2016 at 9:57 PM | 01/14/2016 09:34:06 AM |
| 5 | RW\u0026amp;CO. <… | EXTRA 50% off sale items | Tue, Jan 12, 2016 at 7:05 AM | 01/12/2016 01:26:39 PM |
| 6 | CHEO Foundation <ne… | 🏃Run for kids with cancer at CHEO! Early bird pricin… | | |
| 7 | Camp Fortune <bulleti… | Offre voyage à Banff pour la relâche scolaire! | Fri, Nov 20, 2015 at 11:30 AM | 11/20/2015 04:45:01 PM |
| 8 | Twitter <info@e.twitter… | Al, what are you watching? | Thu, Nov 19, 2015 at 9:15 AM | 11/19/2015 02:24:20 PM |
| 9 | Air Canada Flight Offer… | Great offers for Germany, Asia and more | Wed, Nov 18, 2015 at 3:06 A… | 11/18/2015 09:15:40 AM |

Go To Page: Showing results 1 - 29 of 29

**Figure 20:** IEF Gmail hits from RAM dump

The live-memory capture shows not only the browser install location, but also the user account in which it was installed (see **Figure 21**).

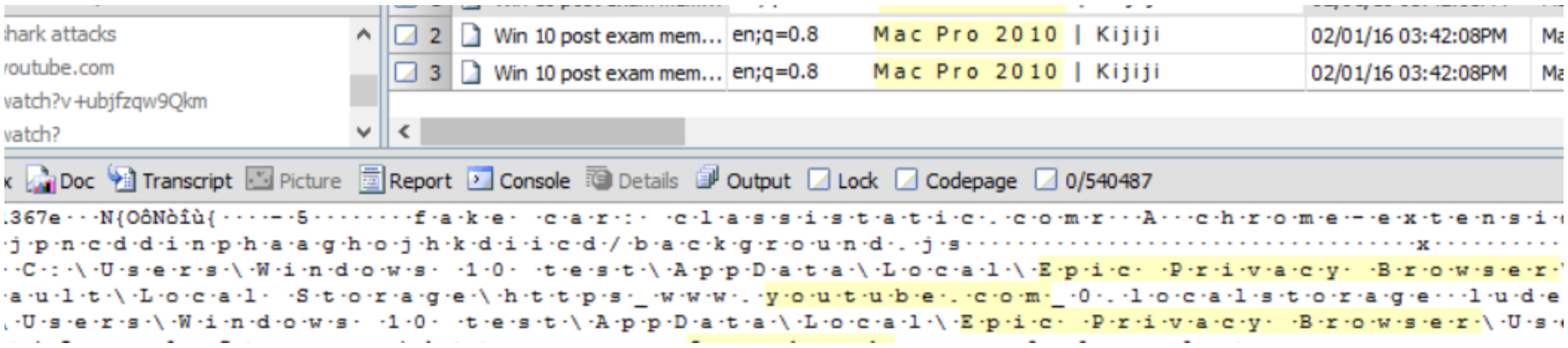

**Figure 21:** Location and user information of Epic Windows 10 RAM dump

Fifty-two thousand hits were recorded from the combined Keyword searches entered in Encase, against the live-memory dumps of Epic queries on both Windows 7 and Windows 10 operating systems. Of the 52,000 hits, approximately 38,000 were recorded from the Windows 10 operating system.

## Discussion

What artefact evidence is produced when the Epic Privacy Browser is installed on the Windows 10 operating system platform? On installation, the application creates a number of documents in the C:\Users\User\AppData\Local\Epic Privacy Browser folder. A default folder is also created that houses data on installation and temporary files and folders used only when the browser is launched. Even though the temporary files are deleted on closure, a great deal of information can be retrieved from both live and post-mortem examination. Registry entries, specific to the user account (SID), are populated and recovered using software such as Registry Viewer. On Windows 7, Epic choses the same location for application installation and, by default, installs the same files and folders as with Windows 10 (with the exception of the *bookmarks.bak* included in Windows 10). Artefact evidence is written to areas such as the pagefile.sys, and little effort is made to delete and overwrite private browsing data.

Another important question is whether all Internet artefact evidence is cleared when the Epic Privacy Browser is closed. Although temporary files and folders within the default folder of the Epic Browser are cleared when the application is closed, the data appears readily available to the forensic examiner, using the standard tools. These remnant traces are similar to those discovered for Browzar (http://www.browzar.com/), another privacy-focused web browser. Upon closing the browsing session, Browzar removed all traces of web browser activity. However, using a combination of forensic tools and techniques, evidence (including pictures, keyword searches, and URLs) was easily recovered in both the memory and in the pagefile (Warren, El-Sheikh & Le-Khac 2017).

Looking at the live-data forensics approach, live-memory capture proves fruitful for the acquisition of Epic artefact evidence. Finding a computer running with the application displayed or minimised on screen would afford the examiner the opportunity to extract the browser 'default' folder in its entirety, thereby capturing all the temporary files and data within. Live-memory dump would also glean a wealth of information, as demonstrated in this study. Acquisition and analysis of the imaged drive has shown to be of benefit from both the Windows 7 and Windows 10 operating systems. Important artefact evidence was found in deleted data files, *pagefile.sys, hiberfil.sys, Ntuser.dat* log files, and unallocated space. It appears that the browser does very little to either overwrite the information or prevent the data's being written to the drive. So in terms of the differences between artefact evidence recovered using the Epic Privacy Browser on Windows 10 and Windows 7 Operating systems, it appears that Windows 7 is far more RAM dependent than its successor; and so far, more evidence was found on the drive. Windows 10 RAM dump produced 80% for the live-memory data from keyword searches. In the case of Browzar forensic analysis, live analysis also proved to contain valuable artefacts: keyword searching, websites visited, and pictures were recovered. In some cases, pictures could not be fully recoverable, but they showed the activities (and their focus) being performed during the browsing session (Warren, El-Sheikh & Le-Khac 2017).

Besides, both *ChromeHistoryView* and *ChromeCacheView* were successful in presenting data acquired from the Epic browser default folder. This was expected since both Google Chrome and Epic Privacy Browser hail from the Chromium source code.

## Conclusion and Future Work

In this paper, the authors presented the forensic acquisition and analysis of the Epic Privacy Browser on Windows 7 and Windows 10. The Epic Privacy Browser prides itself on protecting the user's privacy when online and purports to clear all traces of browsing history on closure. The files and folders created on a temporary basis do get deleted at the end of a browsing session, but the information is still readily available to any forensic examiner using the standard tools. Windows 10 live-memory data produced the bulk of Epic artefact evidence in this operating system, although data was also written to the drive in the areas listed above. The results of this research are useful to, and may be referenced by, forensic experts involved in investigations concerning web activity and for those seeking advanced techniques and methods for recovering, parsing and analysing web-browser-specific data.

## References

1. Aggarwal, G., Bursztein, E., Jackson, C., & Boneh, D. (2010, August). An analysis of private browsing modes in modern browsers. In Proceedings of the 19th USENIX conference on Security (pp. 6-6). USENIX Association.

2. Bissias, G., Levine, B., Liberatore, M., Lynn, B., Moore, J., Wallach, H., & Wolak, J. (2016). Characterization of contact offenders and child exploitation material trafficking on five peer-to-peer networks. Child abuse & neglect, 52, 185-199.

3. Choi, J. H., Lee, K. G., Park, J., Lee, C., & Lee, S. (2012). Analysis framework to detect artifacts of portable web browser. Information Technology Convergence, Secure and Trust Computing, and Data Management, 207-214.

4. Conlan, K., Baggili, I., & Breitinger, F. (2016). Anti-forensics: Furthering digital forensic science through a new extended, granular taxonomy. Digital Investigation, 18, S66-S75.

5. Connolly, M., Niebuhr, J., & Bernnat, R. (2011). Limiting the Impact of Data Breach: The Case of the Sony Playstation Network. Booz & Company, viewed 1 December 2017. <https://www.strategyand.pwc.com/media/file/Limiting-the-impact-of-data-breaches.pdf>

6. Epic Privacy Browser Homepage 2017, viewed 1 December 2017 <https://www.epicbrowser.com/>

7. Farina, J., Kechadi, M., & Scanlon, M. (2015). Project Maelstrom: Forensic Analysis of the BitTorrent-Powered Browser. Journal of Digital Forensics, Security and Law, 10(4), 10.

8. Farina, J., Scanlon, M., Le-Khac, N. A., & Kechadi, M. T. (2015, August). Overview of the forensic investigation of cloud services. In Availability, Reliability and Security (ARES), 2015 10th International Conference on (pp. 556-565). IEEE.

9. Fox-Brewster, T. (2015). Ashley Madison Breach Could Expose Privates of 37 Million Cheaters. *Forbes*, 20 July 2015, viewed 12 September 2017, <https://www.forbes.com/sites/thomasbrewster/2015/07/20/ashley-madison-attack/#151613d55f48>.

10. Gabet, R. M. (2016). A Comparative Forensic Analysis of Privacy Enhanced Web Browsers. MS Thesis, Purdue University, West Lafayette, IN, USA.

11. Hedberg A. (2013). The privacy of private browsing. Technical Report, Tufts University, MA, USA.

12. Hitchcock, B., Le-Khac, N-A., & Scanlon, M. (2016). Tiered forensic methodology model for Digital Field Triage by non-digital evidence specialists. Digital investigation, 16, S75-S85.

13. Khanikekar, S.K. (2010). Web Forensics. Graduate Thesis, Texas A&M University, College Station, TX, USA, <http://sci.tamucc.edu/~cams/projects/345.pdf>.

14. Marrington, A., Baggili, I., Al Ismail, T., & Al Kaf, A. (2012, December). Portable web browser forensics: A forensic examination of the privacy benefits of portable web browsers. In Computer Systems and Industrial Informatics (ICCSII), 2012 International Conference on (pp. 1-6). IEEE.

15. Reed, A, Scanlon, M and Le-Khac, N-A 2017, 'Forensic Analysis of Epic Privacy Browser on Windows Operating Systems', *Proceedings of the 16th European Conference on Cyber Warfare and Security (ECCWS 2017),* Dublin, Ireland.

16. Regshot (2016), 4 November, viewed 13 September 2017, <https://sourceforge.net/projects/regshot/>.

17. Rubenking, N, (2014). Epic Privacy Browser. *PC Magazine*, 6 January 2014, accessed 12 September 2017, <http://in.pcmag.com/epic-privacy-browser/70902/review/epic-privacy-browser>.

18. Scanlon, M., Farina, J., & Kechadi, M. T. (2015). Network investigation methodology for BitTorrent Sync: A Peer-to-Peer based file synchronisation service. Computers & Security, 54, 27-43.

19. Sgaras, C., Kechadi, M. T., & Le-Khac, N. A. (2015). Forensics acquisition and analysis of instant messaging and VoIP applications. In Computational Forensics (pp. 188-199). Springer, Cham.

20. Sha, M. M., Manesh, T., & El-Atty, S. M. A. (2016). VoIP Forensic Analyzer. The International Journal of Advanced Computer Science and Applications (IJACSA), 7, 106-116.

21. Toxen, B 2014, 'The NSA and Snowden: Securing the all-seeing eye', *Communications of the ACM*, vol. 57, no. 5, pp. 44-51.

22. Van Dongen, WS 2007, 'Forensic artefacts left by Windows Live Messenger 8.0', *Digital Investigation: The International Journal of Digital Forensics & Incident Response*, vol. 4, no. 2, pp.73-87, DOI: 10.1016/j.diin.2007.06.019.

23. Voorst VR, Kechadi, M-T., & Le-Khac, N-A 2016, 'Forensic acquisition of IMVU: A case study', *Journal of Digital Forensics, Security and Law*, vol. 10, no. 4, pp.69-77.

24. Warren C, El-Sheikh E & Le-Khac, N-A 2017, 'Privacy preserving Internet browsers: Forensic analysis of Browzar', *Computer and network security essentials*, ed. K Daimi, Springer, Cham. pp. 369-88, DOI: https://doi.org/10.1007/978-3-319-58424-9_21.